# Photonic Hybrid Precoding for Millimeter Wave RoF Systems

Yahia Alghorani, *Member, IEEE*

*Abstract*—Large propagation path loss and limited scattering of millimeter wave (mmwave) channels create new challenges for physical layer signal processing. Hence, in this article, we propose a photonic hybrid precoding model for mmwave radio-over-fiber systems that overcomes hardware constraints on radio frequency precoding. With help of silicon photonic integrated circuits (Si-PICs), the photonic precoding strategy combats the large losses of mmwave and mitigates inter-user interference. Simulation results show that the proposed model can offer significant coding gains over conventional precoding systems.

*Index Terms*—Hybrid precoding, radio-over-fiber systems, millimeter wave communications, multiuser systems.

## I. INTRODUCTION

Hybrid analog-digital precoding transceivers have recently attracted considerable attention for millimeter wave (mm-wave) multiple-input multiple output (MIMO) systems [1],[2]. Such kind of transceivers only require a small number of radio frequency (RF) chains (i.e., signal mixers, highly linear power amplifiers, filters, and high-resolution digital-to-analog (DAC) converters) to provide high precoding (beamforming) gains and combat large path loss of mmwave signals [3]. Due to the high-power consumption of mmwave circuit components and nonlinear electric phase shifters, plus the feedback overhead of complete knowledge of channel state at the transmitter, the conventional hybrid precoding systems performance becomes limited when multiuser interferences are considered in small-cells environments. Thus, radio-over-fiber (RoF) technology is becoming a promising solution to overcome the RF hardware limitations and to increase network capacity [4]. The use of RoF technologies (i.e., elastic radio-optical network (ERON)) for 5G wireless access can bring connection ubiquity, mobility together with high bandwidth and low-latency [5]. As Fig.1 illustrates, ERON topology delivers elastic networking across centralized wireless communications systems, by deploying EON in wide-area and metro-area network (WAN and MAN), and by introducing microwave-optical signal processing at local area network (LAN). The target of ERON structure is to localize digital base-band units (BBUs) in a central office to reduce the complexity of conventional cell sites and deploy a large-scale small-cell system [6]. Since the BBU is shifted to the central office, RF signals are generated at the BBU and transmitted to remote access units (RAUs) through optical fibers. Hence, low-loss wideband (75-110 GHz) channels can be transported over large distances with an increase in system capacity.

This work was supported in part by ARL under grant W911NF-14-2-0114 and by NSF under grant NeTS 1619173.
The author is with the Department of Electrical and Computer Engineering, University of California at Davis, Davis, California 95616 USA (email: yalghorani@ucdavis.edu).

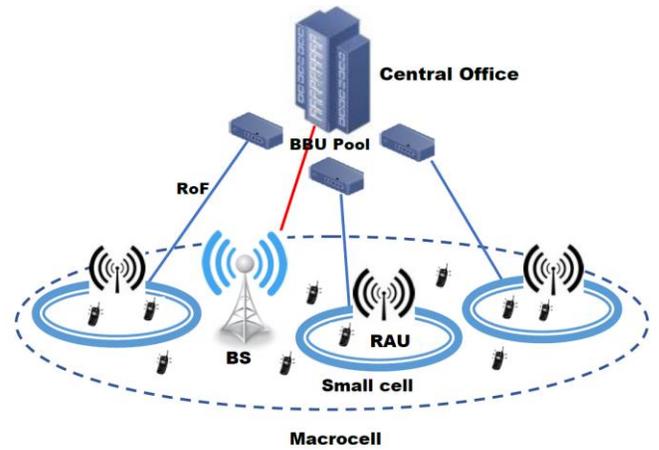

Fig.1. Illustration of ERON architecture

The key advantages of ERON is 1) achieving low-cost, low-noise, low-power, high-fidelity RF-photonic signal processing via silicon photonic integrated circuits (Si-PICs) [7], while replacing the power-hungry mmwave electronic circuitry and nonlinear electric phase shifters. 2) Unlike the baseband-over-fiber (BoF) approach that only supports one frequency band at a time, the RoF approach provides multiple bands multiple carriers signals so that multi-operator can coexist in a shared infrastructure without interference [4]. However, most existing works focus on the hybrid precoding with conventional digital baseband transmission (see e.g., [1], [2]). Hence, in this study, we propose a new RoF system architecture (hybrid precoding based) that bring mobility and interference suppression in multiuser mobile networks.

*Notation*

Boldface symbols denote matrices ($\mathbf{A}$) and vectors ($\mathbf{a}$). The superscripts $(.)^T$ and $(.)^*$ stand for transpose and conjugate, transpose operators, respectively. $(.)^{-1}$ denotes the matrix inverse. The determinant and trace operators are denoted by $\det(.)$ and $\text{Tr}(.)$. $\mathbf{I}$ denotes the identity matrix and $\text{diag}(.)$ denotes a diagonal matrix, $|.|$ is the magnitude of a scalar, and $\|.\|_F$ stands for the Frobenius norm. The Moore-Penrose

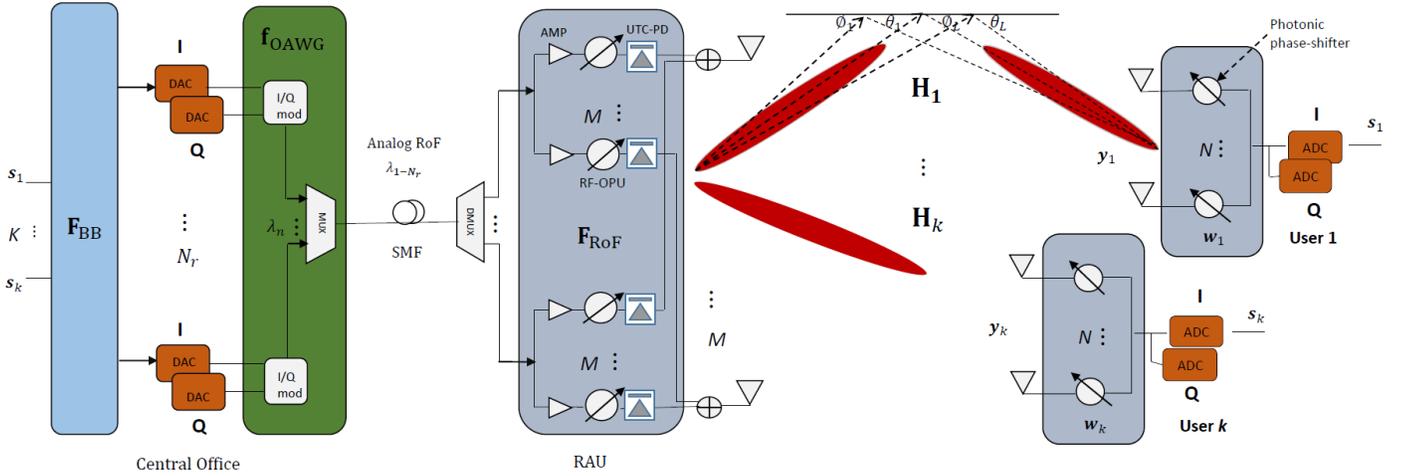

Fig.2. Photonic hybrid precoding structure for mmwave multiuser MIMO systems

pseudoinverse operator is denoted by $(.)^\dagger$.

## II. SYSTEM AND CHANNEL MODEL

We consider a multiuser MIMO system equipped with $M$ transmit and $N$ receive antennas at mmwave. The single-cell system is operating in the ERON and consisting of a digital baseband precoder, $\mathbf{F}_{BB} \in \mathcal{C}^{N_r \times K}$, located at the central office, and an analog photonic precoder (beamformer), $\mathbf{F}_{RoF} \in \mathcal{C}^{M \times N_r}$, at the transmitter (RAU), while the receiver uses only an analog photonic combiner, $\mathbf{w}_k \in \mathcal{C}^{N \times 1}$, as shown in Fig.2. At the output of the baseband precoder, the data streams for $K$ users are carried on $N_r$ optical carriers and precoded into a set of $N_r$ RF chains (DACs) which is subject to a constraint $K \leq N_r \leq M$. Here, each DAC is connected to a dynamic optical arbitrary waveform generator (OAWG) [8], which is, in turn, used to generate spectral slices through an in-phase and quadrature (IQ) optical intensity modulators array, $\mathbf{f}_{OAWG} \in \mathcal{C}^{N_r \times 1}$. Correspondingly, the IQ modulators carry the data streams over RF signals which then transported by a single mode optical fiber (SMF). At the transmitter, the photonic beamformer employs Si-PICs consisting of an optical power splitter, an optical amplifiers array, RF-optical processing units (RF-OPU) for phase/amplitude tuning, and uni-traveling carrier photo-detectors (UTC-PD) to produce the mmwave signals [9]. Given the system model, our main target focuses on adaptive beam steering and interference canceling through the Si-PICs.

We adopt a narrowband-block fading propagation channel model with a constant mmwave channel matrix, $\mathbf{H}_k \in \mathcal{C}^{N \times M}$, between the RAU and the user. Unlike the conventional RF systems; in order to achieve a high multiplexing gain with low-complexity signal processing, the RAU can rely on the characteristics of a multi-carrier optical transmission when accurate channel knowledge is limited due to estimation errors and quantization. However, accurate and timely acquisition of channel state information (CSI) is challenging, especially in inter-user interference and high mobility scenarios [10]. Thus, our proposed scheme can offer a significant gain in network throughput with the low-cost hardware implementation.

With Zero-forcing (ZF) precoding at the RAU, the received signal at user $k$, can be written as

$$y_k = \sqrt{\rho} \mathbf{w}_k^H \mathbf{H}_k \mathbf{x}_k + \sqrt{\rho} \sum_{m \neq k}^{K} \mathbf{w}_k^H \mathbf{H}_k \mathbf{x}_m + \mathbf{w}_k^H \mathbf{n}_k \quad (1)$$

where $\rho = P_s / K$ is the transmit signal power per user, $\mathbf{x}_k = \mathbf{F}_{RoF} \mathbf{f}_{OAWG} \mathbf{F}_{BB}^{(k)} \mathbf{s}_k$ is the signal vector transmitted by the RAU, $\mathbf{s}_k \in \mathcal{C}^{N_s \times 1}$ is the $k$-th transmit symbol vector, and $\mathbf{n}_k \in \mathcal{C}^{N \times 1}$ is the additive white Gaussian noise (AWGN) vector with zero mean and variance $N_o$. The normalized total transmit power constraint is given by $\|\mathbf{F}_{RoF} \mathbf{f}_{OAWG} \mathbf{F}_{BB}\|_F^2 = N_r K$. Therefore, the received signal-to-interference-and-noise ratio (SINR) at $k$ user becomes

$$\text{SINR}_k = \frac{\rho \left| \mathbf{w}_k^H \mathbf{H}_k \mathbf{F}_{RoF} \mathbf{f}_{OAWG} \mathbf{F}_{BB}^{(k)} \right|^2}{\rho \sum_{m \neq k}^{K} \left| \mathbf{w}_k^H \mathbf{H}_k \mathbf{F}_{RoF} \mathbf{f}_{OAWG} \mathbf{F}_{BB}^{(m)} \right|^2 + \|\mathbf{w}_k\|^2 N_o} \quad (2)$$

The corresponding link spectral efficiency is $\log_2(1 + \text{SINR}_k)$ bits/sec/Hz, where the digital precoder is selected by [11]

$$\mathbf{F}_{BB}^{(k)} = \mathbf{H}_{p(k)}^\dagger = \mathbf{H}_{p(k)}^H \left( \mathbf{H}_{p(k)} \mathbf{H}_{p(k)}^H \right)^{-1} \quad (3)$$

where $\mathbf{H}_{p(k)} = \mathbf{w}_k^H \mathbf{H}_k \mathbf{F}_{RoF} \mathbf{f}_{OAWG}$. The ZF precoding aims at nulling inter-user interference but in return neglects the effect of noise, which means that it works poorly under noise-limited scenarios. In contrast, minimum mean-square error (MMSE) precoding can provide a tradeoff between noise amplification and interference suppression, at the expense of increasing system complexity. This precoder can be expressed as

$$\mathbf{F}_{BB}^{(k)} = \mathbf{H}_{p(k)}^H \left( \mathbf{H}_{p(k)} \mathbf{H}_{p(k)}^H + \frac{\|\mathbf{w}_k\|^2 \mathbf{I}_K}{\text{snr}} \right)^{-1}$$

where the signal-to-noise ratio (SNR) is defined as $\text{snr} = P_s / N_o$. In general, both the linear ZF and MMSE precoders may achieve a full multiplexing gain if the complete CSI is fed back to the transmitter with good accuracy, thus, the photonic beamformers-based RoF systems can compensate the signal power loss caused by imperfect CSI at the transmitter, through spreading multiple data streams over multiple carriers.

Since the mmwave channels have large propagation path loss and limiting scattering, we use a geometric channel model between the RAU and the user, described by [12]

$$\mathbf{H}_k = \sqrt{\frac{MN}{L_k}} \sum_{l=1}^{L_k} \alpha_{k,l} \mathbf{a}_r(\theta_{k,l}) \mathbf{a}_t^H(\emptyset_{k,l}) \quad (4)$$

where $L_k$ is the number of scatters at user $k$, $\alpha_{k,l}$ is the complex gain of path $l$, $\theta_{k,l} \in [0, 2\pi]$ and $\emptyset_{k,l} \in [0, 2\pi]$ are the angle of arrival (AoA) and the angle of departure (AoD), respectively. With uniform linear arrays (ULA), the $k$-th array response vector at the RAU is expressed as

$$\mathbf{a}_t(\emptyset_{k,l}) = \frac{1}{\sqrt{M}}\left[1, e^{j\frac{2\pi}{\lambda_n}d\sin(\emptyset_{k,l})}, \ldots, e^{j(M-1)\frac{2\pi}{\lambda_n}d\sin(\emptyset_{k,l})}\right]^T \quad (5)$$

where $d$ is the distance between antenna elements and $\lambda_n$ is the signal wavelength of the carrier $n$ ($n = 1, 2, \ldots, N_r$). A similar expression can be written for the array response vector at user $k$, $\mathbf{a}_r(\theta_{k,l})$.

Despite the silicon photonic lattice filter array is used to apply appropriate amplitude and phase weights, we assume that all phase shifters have the same amplitude fixed at $1/\sqrt{M}$.

### III. PROBLEM FORMULATION

Because of the high bandwidth of mmwave, conventional RF beamformers based on electric phase shifters cause phase-shifter loss, high noise power and nonlinearity, which results in beam squint (steering error) across frequency. In addition, the use of large channel bandwidth leads to high noise power and low received SNR, which makes it difficult to implement functions like channel estimation and beam training.

Since mmwave signals occupy a very small portion of the optical bandwidth, linear RF phase shifts are achievable by using a linear photonic filter [13]. Research studies have demonstrated that RF-photonic technologies can perform low relative-intensity-noise lasers, highly linear optical modulators for broadband signals, and highly linear photo-detectors even at high power levels [14], [15]. All these aspects are very important when realizing high-performance RoF links with high SNR values. Thus, our goal in using the broadband photonic beamformer is to maximize the SNR through noise and interference cancellation.

Given the array steering vector in (5), the $k$-th array gain of the photonic beamformer $\mathbf{F}_{\text{RoF}}$ is expressed as

$$|g(x_{k,n})| = \frac{1}{\sqrt{MN_r}} \sum_{n=1}^{N_r} \sum_{m=1}^{M} e^{j(m-1)\frac{2\pi}{\lambda_n}dx_{k,n}} \quad (6)$$

where $x_{k,n} = \xi_n \sin(\phi_{k,n}) - \sin(\phi_{k,o})$ is the effect of beam squint for line-of-sight (LOS) scenarios (i.e., $L = 1$), $\xi_n \in [1 - b_n/2, 1 + b_n/2]$ stands for the ratio of a subcarrier to the carrier $f_n$, $b_n = BW_n/f_n$ is the fractional bandwidth, $\phi_{k,n}$ is the $k$-th AoD for $f_n$, and $\phi_{k,o}$ is the $k$-th beam focus angle, in which the photonic beamformer can achieve the highest array gain. In contrast, the $k$-th array gain of the narrowband (i.e.,

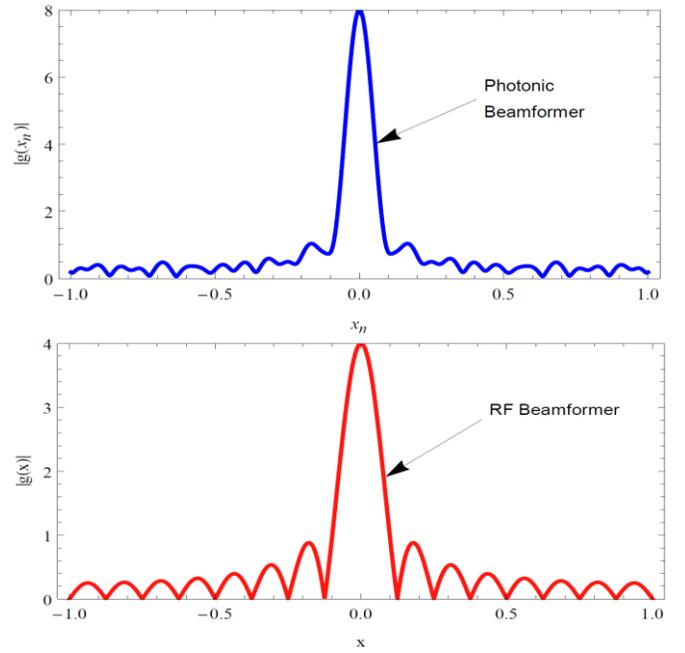

**Fig.3**. Comparison between the array gain of RF and photonic beamformers. A fine beam pattern for each user, generated by 16 antenna elements with an element spacing of $\lambda/2$ for a carrier frequency 28 GHz. The broadband beamformer assigns a set of 4 wavelengths ($\lambda_1 = 10.70$mm, $\lambda_2 = 7.1$mm, $\lambda_3 = 4.99$mm, $\lambda_4 = 4.10$mm) used by each antenna element in the array, while the RF beamformer employs a single-wavelength ($\lambda = 10.70$mm). The maximum array gain ratio is $|g(x_n)|/|g(x)| = \sqrt{N_r}$.

single-carrier) RF beamformer $\mathbf{F}_{\text{RF}}$, is expressed as

$$|g(x_k)| = \frac{1}{\sqrt{M}} \sum_{m=1}^{M} e^{j(m-1)\frac{2\pi}{\lambda}dx_k} \quad (7)$$

From (6) and (7), a phase shift causes different frequency components to be steered to different angles. Meaning that the beam steering errors increase if the AoD/AoA gets away from the desired beam steering angle $\phi_{k,o}$. To get further insight into the performance of the photonic beamformer, we need to find an asymptotic expression for the $k$-th array gain. Assuming the photonic phase shifters have an identical gain for any optical carrier $\lambda_n$, then using the inequality $e^{ax} \geq (1 + x)^a \geq 1 + ax$ when $x \to 0$, the photonic array gain can be bounded as $|g(x_k)| \geq \sqrt{N_r/M}(1 + j(M - 1)2\pi\lambda^{-1}dx_k)$. Which concludes that the beam squint errors generated by the coefficient $\xi_n$, can effectively be minimized when the number of wavelengths $\lambda_n$ increases. Fig.3 illustrates an example fine beam pattern for both the photonic and RF beamformers, where we assume that the phase-shifter array is steered to the desired angle for the carrier frequency (i.e., $\xi_n = 1$). The results show that the photonic beamformer can provide a narrow directional beam with a higher gain of signal power, which indicates that the photonic beamformer can combat Doppler shift effects and tackle the low SNR and inter-user interference at mmwave.

### IV. PHOTONIC HYBRID PRECODING FOR MM WAVE MULTI-USER MIMO SYSTEMS

To simplify the analysis of the photonic hybrid precoding, we focus on the case that each user is equipped with a single-antenna (i.e., $N = 1$). Hence, the received signal vector, $\mathbf{y} \in \mathcal{C}^{K \times 1}$, is given by

$$\mathbf{y} = \sqrt{\rho}\mathbf{H}\mathbf{x} + \mathbf{n} \qquad (8)$$

where $\mathbf{H} \in \mathcal{C}^{K \times 1}$ is the channel response for $K$ users, which is subject to the constraint $\|\mathbf{H}\|^2 = \mathrm{Tr}(\mathbf{H}\mathbf{H}^H) = \sum_{k,l} M|\alpha_{k,l}|^2/LK$, $\mathbf{x} = \mathbf{F}_{\mathrm{RoF}}\mathbf{f}_{\mathrm{OAWG}}\mathbf{F}_{\mathrm{BB}}\mathbf{s}$ is the transmit signal vector ($\mathbf{s} \in \mathcal{C}^{K \times 1}$, $\mathbf{x} \in \mathcal{C}^{M \times 1}$), and $\mathbf{n} \in \mathcal{C}^{K \times 1}$ represents the noise and interference.

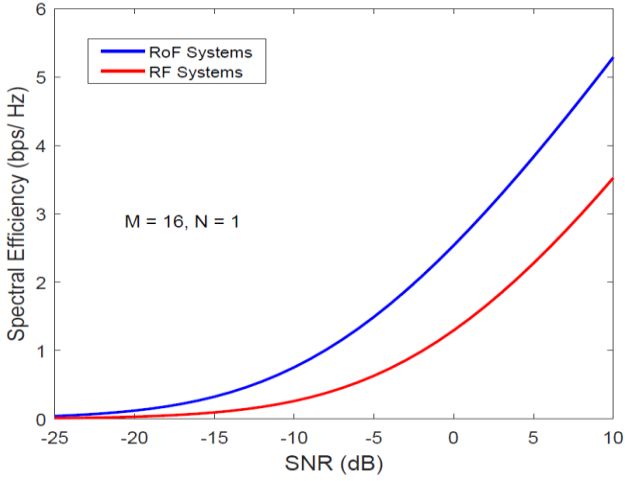
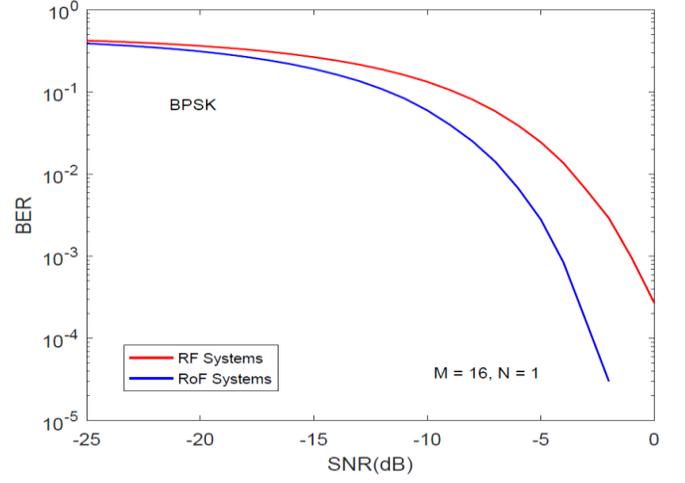

**Fig.4.** Spectral efficiency and error probability comparison of conventional RF and RoF systems. Both the considered schemes are achieved by using the hybrid precoding approach (ZF precoding based). Monto-Carlo simulation is implemented at $K = N_r = 3$, $\lambda_{1-N_r} = d/2$, $L = 1$ (LOS paths), and $10^5$ channel realizations.

*A. Achievable Spectral Efficiency*

In our analysis, we focus on the downlink transmission. Plus, we assume that the channel $\mathbf{H}$ is known to the transmitter. The transmit symbols follow a Gaussian distribution with Rayleigh fading, in which achievable spectral efficiency is expressed as

$$R = \log\det\left(\mathbf{I_K} + \frac{\rho}{N_o}\mathbf{H}\mathbf{F}_{\mathrm{RoF}}\mathbf{f}_{\mathrm{OAWG}}\mathbf{F}_{\mathrm{BB}}\mathbf{F}_{\mathrm{BB}}^H\mathbf{f}_{\mathrm{OAWG}}^H\mathbf{F}_{\mathrm{RoF}}^H\mathbf{H}^H\right) \quad (9)$$

To maximize the spectral efficiency in (9), we assume that the columns of the digital precoding matrix are mutually orthogonal, where the interference between the data streams is canceled, namely

$$\mathbf{F}_{\mathrm{BB}}^H\mathbf{F}_{\mathrm{BB}} = \sigma\mathbf{I}_K \qquad (10)$$

Proceeding with the design of the photonic analog precoder, $\mathbf{F}_{\mathrm{RoF}}\mathbf{f}_{\mathrm{OAWG}}$, and the digital precoder, $\mathbf{F}_{\mathrm{BB}}$, the corresponding optimization problem is given by

$$\min_{\mathbf{F}_{\mathrm{RoF}},\mathbf{f}_{\mathrm{OAWG}},\mathbf{F}_{\mathrm{BB}}} \|\mathbf{F}_{\mathrm{opt}} - \mathbf{F}_{\mathrm{RoF}}\mathbf{f}_{\mathrm{OAWG}}\mathbf{F}_{\mathrm{BB}}\|_F^2$$

$$\text{subject to } \begin{cases} \mathbf{F}_{\mathrm{RoF}} \in \mathcal{F}_{\mathrm{RoF}} \\ \mathbf{f}_{\mathrm{OAWG}} \in \mathcal{F}_{\mathrm{OAWG}} \\ \mathbf{F}_{\mathrm{BB}}^H\mathbf{F}_{\mathrm{BB}} = \sigma\mathbf{I}_K \end{cases} \quad (11)$$

where $\mathbf{F}_{\mathrm{opt}}$ stands for the optimal digital precoding (i.e, $N_r = M$) [1], and $\sigma$ is an optimization factor selected to minimize the objective function $\|\mathbf{F}_{\mathrm{opt}} - \mathbf{F}_{\mathrm{RoF}}\mathbf{F}_{\mathrm{OAWG}}\mathbf{F}_{\mathrm{BB}}\|_F^2$.

The photonic beamformer, $\mathbf{F}_{\mathrm{RoF}}$, and the optical modulator array, $\mathbf{f}_{\mathrm{OAWG}}$, are chosen to maximize the mutual information between the transmitter and receiver, under a constraint of $\|\mathbf{F}_{\mathrm{RoF}}\mathbf{F}_{\mathrm{OAWG}}\|_F^2 = K$. Using the identity $\|A - B\| \geq |\|A\| - \|B\||$, the objective function $\|\mathbf{F}_{\mathrm{opt}} - \mathbf{F}_{\mathrm{RoF}}\mathbf{F}_{\mathrm{OAWG}}\mathbf{F}_{\mathrm{BB}}\|_F^2$ can be further simplified as

$$\|\mathbf{F}_{\mathrm{opt}} - \mathbf{F}_{\mathrm{RoF}}\mathbf{F}_{\mathrm{OAWG}}\mathbf{F}_{\mathrm{BB}}\|_F^2 \geq \left|\|\mathbf{F}_{\mathrm{opt}}\| - \|\mathbf{F}_{\mathrm{RoF}}\mathbf{F}_{\mathrm{OAWG}}\mathbf{F}_{\mathrm{BB}}\|\right|^2$$

$$\geq \left|\sqrt{N_r K} - \sqrt{\sigma K}\right|^2 \quad (12)$$

Thus, when $\sigma = N_r$, the objective function has the minimum value, given by $\|\mathbf{F}_{\mathrm{opt}}\|^2 - \sigma$, resulting in achievable spectral efficiency expressed as

$$R \leq \sum_k \sum_l \log_2\left(1 + \frac{\mathrm{snr}}{LK^2}MN_r(K-1)|\alpha_{k,l}|^2\right) \text{ bits/s/Hz} \quad (13)$$

Notice here that the spatial multiplexing gain (degrees of freedom) extracted from the channel $\mathbf{H}$, is $LK$. In addition, the worst-case for the photonic hybrid precoding is when $N_r = K$, and the best-case is when $N_r = M$ (like in the case of the optimal digital precoding). Since the photonic beamforming improves the signal power of users, there is no need to implement more RF chains at transceivers.

*Special cases:* A user at the edge of a cell may experience a low SNR and interference. Therefore, using the approximation $\log_2(1 + x) \approx x \log_2 e$ for $x$ small, we obtain

$$R_{\mathrm{snr}\to 0}^{(k)} \leq \frac{MN_r}{K^2}(K-1)\,\mathrm{snr}\log_2 e \text{ bits/s/Hz} \quad (14)$$

At low SNR, each user yields a power gain of $MN_r(K-1)/K^2$. Note that when the number of scatters $L$ increases, the power gain does not increase. On the other hand, when increasing the number of transmit antennas and/or the number of optical carriers, the power gain is boosted by the photonic beamformer, resulting in the spectral efficiency enhancement. Hence, sending the users' data streams over multi-carrier frequencies makes the RoF systems maximize the received SNR even if the transmitter has limited channel knowledge. This advantage will allow the photonic hybrid precoding to overcome the problem of inter-symbol interference and inter-user interference for conventional multiuser MIMO systems at mmwave [16]. While this problem was practically solved for wideband single-user systems through orthogonal frequency division multiplexing (OFDM), there are still challenges of implementing the multiuser precoding at mmwave [2].

With a short wavelength of mmwave, it is possible to pack a large number of antennas into the photonic beamformer to overcome the high path loss. For multiuser MIMO systems, the number of transmit antennas greatly exceeds the number of users (i.e, $M \gg K, M \to \infty$), where the row vectors of the channel $\mathbf{H}$ are asymptotically orthogonal. In this case, we have

$$\frac{\mathbf{H}\mathbf{H}^H}{M} \approx \mathbf{I}_K \qquad (15)$$

Assuming $\alpha_{k,l}$ ($l = 1, 2, ..., L$) are independent and identical distributed (i.i.d) $\mathcal{CN}(0, 1)$ random variables with LOS path to the user, the achievable spectral efficiency (9) is approximated as

$$R^{(k)} \approx \log_2(1 + \text{snr}MN_r) \quad \text{bits/s/Hz} \qquad (16)$$

As we notice in (16), massive MIMO systems with multi-carrier transmission can greatly improve the power gain up to $MN_r$, which treats any low SNR and interference. Massive MIMO systems have also the capability of providing steerable beams to compensate large path-loss of mmwave. However, increasing antenna elements at a microcell base station can create narrower beams and higher gains but may increase the channel estimation errors and pilot contamination [16].

In conventional multi-cell massive MIMO systems, users in neighboring cells may use non-orthogonal pilots because the number of pilot sequences is smaller than the number of users, which results in directed inter-cell interference and pilot contamination. Since RoF systems deploy a large-scale small cell network, a small number of orthogonal pilot sequences is required to serve a small number of users per cell. Thus, using massive MIMO systems in small cell networks can completely remove the effect of pilot contamination and maximize the spectral efficiency [17].

*B. Average Bit Error Rate*

Assuming each user employs a maximum likelihood (ML) detector, the bit error rate (BER) for binary phase-shift keying (BPSK) modulation is expressed as $P_b^{(k)} \geq Q(\sqrt{2\gamma_k})$, where $\gamma_k = \sum_l \text{snr}MN_r(K-1)|\alpha_{k,l}|^2 / LK^2$ is the effective SNR at user $k$.

Following the derivation in [18, eq.(3.44)] and assuming the complex gains $\alpha_{k,l}$ are i.i.d $\mathcal{CN}(0, 1)$ random variables, the average BER for user $k$ can be simplified in the high SNR regime (i.e., $\text{snr} \gg 1$) as

$$\bar{P}_b^{(k)} \leq \left(\text{snr}\frac{MN_r(K-1)}{LK^2}\right)^{-L} \qquad (18)$$

Notice that the diversity order for the underlying scheme is $d = L$, whereas the coding gain is $r = MN_r(K-1)/LK^2$. Fig 4. plots both the spectral efficiency and the bit error probability of BPSK modulation for hybrid precoding-based RoF MIMO systems. The performance is evaluated in single-path mmwave channels (where we assume the number of scatters is very limited). Multiplexed data streams are sent from an RAU with $M = 16$ antennas to a set of users ($K = 3$), each having a single antenna. In addition, We assume that a zero-forcing precoder is employed at the central office and the number of RF chains and optical carriers, is equal to the number of users ($N_r = K$). While we used a mmwave channel model (4) to simplify our analysis; in our simulation, we perform the Saleh-Valenzuela model with a single propagation cluster, in which the RAU has a uniform square planar array (USPA) with $\sqrt{M} \times \sqrt{M}$ antenna elements [19]. The azimuth and elevation of AoAs/AoDs are uniformly distributed in $[0, 2\pi]$ and $[-\frac{\pi}{2}, \frac{\pi}{2}]$, respectively.

Monto-Carlo simulation is implemented based on (2) and compared with that of conventional RF systems (where the optical intensity modulator array $\mathbf{f}_{\text{OAWG}}$ is omitted and the RF beamformer $\mathbf{F}_{\text{RF}}$ operates a single-carrier). The results show that the RoF systems bring an extra gain over conventional RF systems; which is expected since the photonic beamformer offers a higher power gain than the RF beamformer. This can be deduced directly from (14) and (18), the power and coding gain ratio of the RoF systems to the RF systems is $N_r$. Thus, the photonic hybrid precoders can achieve better interference mitigation in multiuser MIMO systems compared with the conventional hybrid precoders. These findings indicate that: 1) the RoF systems can provide higher data streams per user, resulting in higher data transmission, 2) they can achieve a significant power gain for low SNR and interference scenarios without performing power allocation techniques to maximize the link spectral efficiency, and 3) multi-carrier transmission implies that; multiple broadband signals can be transmitted to multiple users, in such a way that the system dynamically maximizes the coding gain and minimizes the number of RF chains.

## V. CONCLUSION

In this paper, we proposed a new hybrid precoding structure for RoF systems. The system architecture considers both the flexibility and the robustness to RF precoders inaccuracy. The photonic hybrid precoding systems aims to overcome the large propagation loss and reduce channel estimation cost for mm wave. Our proposed model addressed the problem of low SNR and interference in small cell scenarios. The simulation results clearly demonstrated that the RoF systems can offer a higher precoding gain than the RF systems when the numbers of RF chains and users are equal. The photonic hybrid precoder can efficiently overcome multiuser precoding design problems at mmwave, by enabling the broadband beamformer to increase the power gain and mitigate inter-user interference.